\documentclass[superscriptaddress,twocolumn,showpacs,preprintnumbers,amsmath,amssymb, prx, reprint, longbibliography, showkeys]{revtex4-1}
\usepackage{amsmath}
\usepackage{stmaryrd}
\usepackage{txfonts}
\usepackage{amssymb}
\usepackage{mathrsfs}
\usepackage{graphicx}
\usepackage{dcolumn}
\usepackage{bm}
\usepackage{epsfig}
\usepackage{color}
\usepackage{pbox}
\usepackage[colorlinks=true, linkcolor=blue, citecolor=blue, urlcolor=blue]{hyperref} 
\begin{document}
\preprint{\href{https://doi.org/10.1103/PhysRevLett.120.077202}{S.-Z. Lin and C. D. Batista, Phys. Rev. Lett. {\bf 120}, 077202 (2018).}}

\title{Face centered cubic and hexagonal close packed skyrmion crystals in centro-symmetric magnets}
\author{Shi-Zeng Lin}
\email{szl@lanl.gov}
\affiliation{Theoretical Division, T-4 and CNLS, Los Alamos National Laboratory, Los Alamos, New Mexico 87545, USA}

\author{Cristian D. Batista}
\affiliation{Department of Physics and Astronomy, The University of Tennessee, Knoxville, Tennessee
9 37996, USA}
\affiliation{Quantum Condensed Matter Division and Shull-Wollan Center, Oak Ridge National Laboratory, Oak Ridge, Tennessee 37831, USA}

\begin{abstract}
Skyrmions are disk-like objects that typically form  triangular crystals in two dimensional systems.  This situation is analogous to  the so-called ``pancake vortices" of  quasi-two dimensional superconductors. The way in which skyrmion disks or \emph{pancake skyrmions} pile up in layered centro-symmetric materials is dictated by  the inter-layer exchange.
Unbiased Monte Carlo simulations and simple stabilization arguments reveal  face centered cubic and hexagonal close packed skyrmion crystals for different choices of the inter-layer exchange, in addition to the conventional triangular crystal of  skyrmion lines. Moreover, an inhomogeneous current induces  sliding motion of  pancake skyrmions, indicating that they behave as effective mesoscale  particles.
\end{abstract}
\date{\today}
\maketitle

Magnetic skyrmions are swirling spin textures, which have been recently discovered in magnets without inversion symmetry~\cite{Muhlbauer2009, Yu2010a}. Protected by their nontrivial topology, skyrmions are robust against small perturbations and can be driven by various external stimuli~\cite{Everschor12,White2012,Liu2013b,White2014,Kong2013,Lin2014PRL,Mochizuki2014,Jonietz2010,Yu2012,Schulz2012}. Because of their compact size, their high mobility and the possibility of creating or destroying them with electric currents, skyrmions are regarded as promising candidates for applications in memory devices \cite{Fert2013,nagaosa_topological_2013,jiang_skyrmions_2017,fert_magnetic_2017}. It is known by now that skyrmions are rather ubiquitous topological magnetic strcutures because they have been observed in several classes of magnetic materials {\it without inversion symmetry}, including metals \cite{Muhlbauer2009,Yu2010a}, semiconductors \cite{Yu2011}, insulators and multiferroics \cite{Adams2012,Seki2012}. In bulk,   skyrmion structures typically appear as triangular crystals of straight lines parallel to ${\bm H}$. This phenomenon is analogous  to the Abrikosov vortex lattice of type II superconductors. However, the skyrmion crystal (SC) phase of non-centrosymmetric  magnets is only a small pocket of the thermodynamic phase diagram. In thin films, skyrmions become pancake-like objects that still form a triangular lattice. In contrast to the 3D case, this  phase is stable over a wide field and temperature region extending down to $T=0$. 

Different non-centrosymmetric or chiral magnets exhibit similar skyrmion phase diagrams because  the linear skyrmion size, $l_s$, is much bigger than the atomic lattice parameter: $l_s/a >> 1$. This dimensionless ratio is of the order of $J/D$, where $J$ is the ferromagnetic exchange constant and $D$ is the magnitude of a Dzyaloshinskii-Moriya (DM) interaction~\cite{Dzyaloshinsky1958,Moriya60,Moriya60b} arising from the lack of inversion symmetry. The phase diagram is satisfactorily described by an effective continuum model including ferromagnetic exchange, the  DM interaction and the Zeeman term~\cite{Bogdanov89,Bogdanov94,Rosler2006}.

Given that most materials have inversion symmetry, it is relevant to ask if skyrmion crystals can emerge in centro-symmetric magnets. According to  Derrick's theorem~\cite{Derrick1252},  stable topological excitations require the existence of a characteristic length scale. Competing interactions in frustrated magnets can provide this length scale.~\cite{PhysRevLett.108.017206} For instance, helical  magnetic orderings,which are quite ubiquitous in rare-earth magnets~\cite{Yu2012b}, have a characteristic length $2\pi/|{\bm Q}|$ associated to their propagation vector ${\bm Q}$. Indeed, localized spin textures can be stabilized by competing interactions. \cite{PhysRevLett.108.096403,ozawa_vortex_2016,wulferding_domain_2017}   
Centro-symmetric magnets support skyrmions with any sign of the of scalar spin chirality (the skyrmion charge can be positive or negative). Moreover, skyrmions of uniaxial magnets can have  arbitrary \emph{helicity}~\cite{nagaosa_topological_2013}, because of the $U(1)$ symmetry of spin rotation along the field-axis. Besides the additional Goldstone mode of this internal degree of freedom, skyrmions in centro-symmetric magnets have interesting properties not shared by skyrmions of chiral magnets~\cite{leonov_multiply_2015,Lin2016a,PhysRevLett.117.157205,PhysRevLett.118.247203,PhysRevLett.118.147205,PhysRevLett.119.207201,zhang_skyrmion_2017}.

The triangular skyrmion lattice can be regarded as a superposition of three single-${\bm Q}$ magnetic helices with propagation vectors differing by   $ \pm120^\circ$ rotations.  The triple-${\bm Q}$  superposition  forces a spatial modulation of the {\it magnitude} of the magnetic moment, which has an exchange energy cost. In uniaxial 2D magnets with easy-axis anisotropy, this energy  cost can be compensated by an anisotropy energy gain~\cite{leonov_multiply_2015,PhysRevB.93.184413}. Indeed, skyrmions have been recently observed in the centro-symmetric materials~\cite{Yu2012b,wang_centrosymmetric_2016}. 
However, it is unclear how these pancake skyrmions organize in 3D layered magnets.  In this Letter, we study 3D skyrmion crystals emerging in frustrated Heisenberg models on a vertically stacked triangular lattice. We demonstrate that frustration of the inter-layer exchange  leads to  multiple ways of stacking pancake skyrmions along the $c$-axis. Small $Q_z \ll 1$ values lead to triangular SC's of tilted lines relative to the external magnetic field. In contrast, larger values of $Q_z$ produce hexagonal close packed (HCP) and face centered cubic (FCC) crystals of pancake skyrmions. 
  
We consider the spin Hamiltonian 
\begin{align}\label{eq1}
{\cal H} =  \sum \limits_{ i\neq j }{J_{ij}{\bm{S}}_i}\cdot{{\bm{S}}_j}  - H \sum \limits_i {{{S}}_{i,z}} - A \sum \limits_i S_{i,z}^2,
\end{align}
defined on a vertically stacked triangular lattice, which includes an easy-axis anisotropy term ($A>0$). The external magnetic field ${\bm H}$ is assumed to be parallel to the $c$-axis. The intra-layer exchange  includes a nearest-neighbor (NN) ferromagnetic (FM) coupling, $J_1<0$, and a third NN antiferromagnetic (AFM)  interaction $J_3 >0$. The inter-layer exchange is also assumed to be frustrated. We fix the ratio $J_3/J_1=-0.5$, corresponding to a magnitude of the ordering wave vector $Q_{ab}=2\cos^{-1}[(1+\sqrt{1-2J_1/J_3})/4]= 2\pi/5$~\footnote{The in-plane wave vector components are in units of $a^{-1}$, where $a$ is the triangular lattice parameter, while the out of plane component, $Q_z$, is in units of $c^{-1}$, where $c$ is the lattice parameter along the direction perpendicular to the triangular layers. }.   The  2D limit of this Hamiltonian includes a triangular  SC in its phase diagram.~\cite{leonov_multiply_2015,PhysRevB.93.184413}

The exchange interaction in momentum space is
\begin{equation}
\mathcal{H}_{\mathrm{ex}}= \sum \limits_{ i\neq j }{J_{ij}{\bm{S}}_i}\cdot{{\bm{S}}_j}= \sum_{\bm{q}}J(\bm{q})\bm{S}(\bm{q})\cdot\bm{S}(-\bm{q}),
\end{equation}
where $J(\bm{q})$ and $S(\bm{q})$ are the Fourier transform of $J_{ij}$ and $\bm{S}_i$. The ground state is a helix with an ordering wave vector $\bm{Q}$ that minimizes $J(\bm{q})$. The easy-axis anisotropy  distorts the helix by  inducing higher harmonics that  increase  the easy-axis spin component. If $J(\bm{q})$ is minimized by six ordering wave vectors 
 $\pm \bm{Q}_{\mu}$    ($\mu=1,2,3$)  and $\bm{Q}_1+\bm{Q}_2+\bm{Q}_3={\bm 0}$, the  field  ${\bm H}$ favors the formation of skyrmion crystals because it enables an effective interaction of 
the form 
$g  \bm{S}(\bm{Q}_1)\cdot\bm{S}(\bm{Q}_2) \; \bm{S}(\bm{Q}_3)\cdot\bm{S}(\bm{0})$~\cite{Garel_PhysRevB.26.325}. The condition  $\bm{Q}_1+\bm{Q}_2+\bm{Q}_3={\bm 0}$ explains the importance of $C_3$ invariant spin systems for the stabilization of triple-${\bm Q}$ spin structures, such as skyrmion~\cite{leonov_multiply_2015,PhysRevB.93.184413} or vortex crystals~\cite{Kamiya14,Wang15}.

\begin{figure}[t]
\psfig{figure=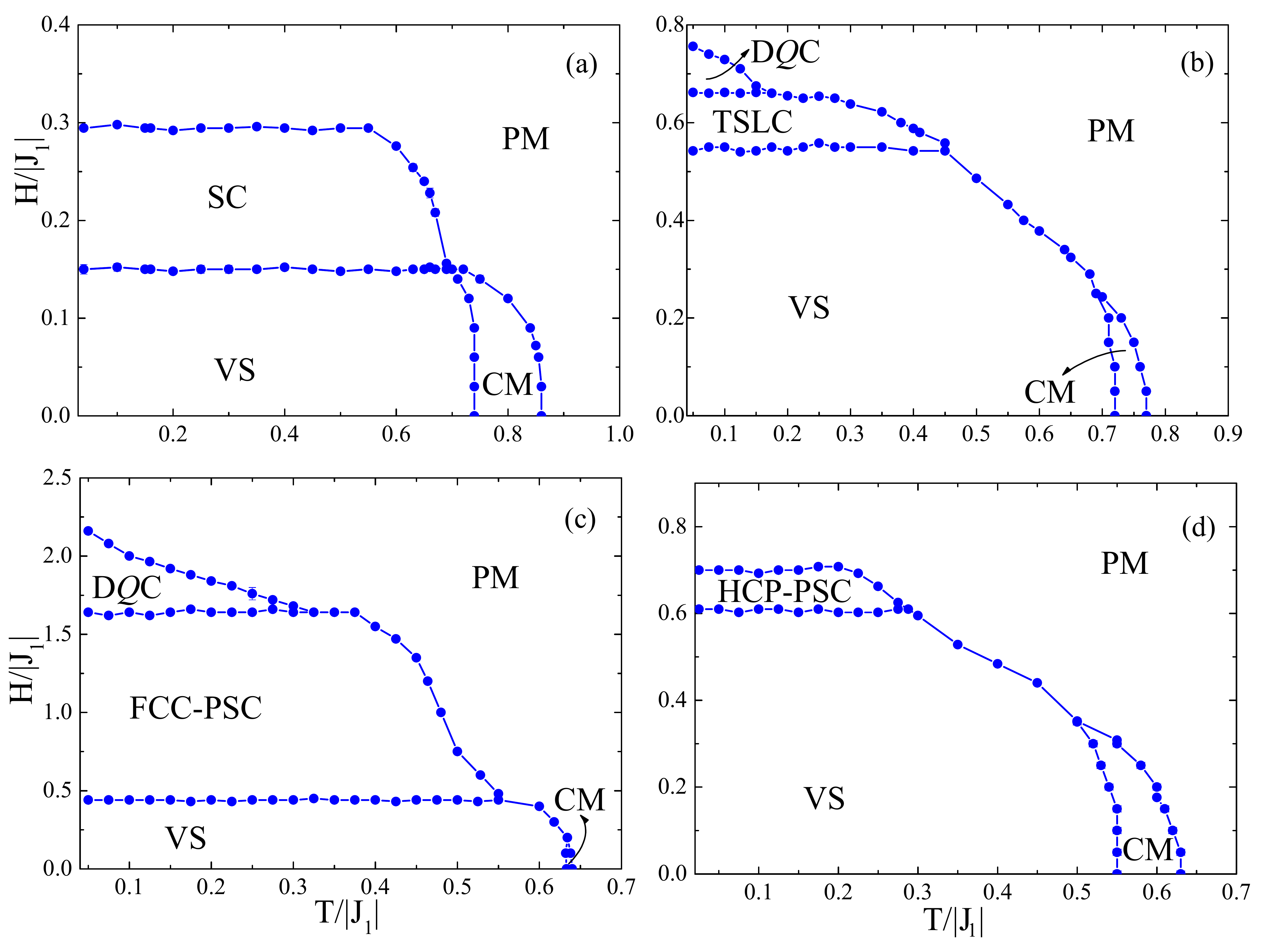,width=\columnwidth}
\caption{(color online) Temperature-magnetic field phase diagram for the Hamiltonian of Eq. \eqref{eq1} with (a) $Q_z=0$, (b) $Q_z=2\pi/5$, (c)  $Q_z=2\pi/3$ and (d)  $Q_z=\pi$, obtained from Monte Carlo simulations. The easy-axis anisotropy is $A = 0.5 |J_1|$. In (a) the interlayer  exchange  is $J^c_1=0.5 J_1$ between adjacent layers and $J^c_2=0$ between NNN layers. In (b),  $J^c_{1}=0.5J_1$ and $J^c_{2}=-0.4045 J^c_1$. In (c),  $J^c_1=-0.5J_1$ and $J^c_2=-0.25 J_1$. In (d), $J^c_{1}=-0.2J_1$ and $J^c_2=0$.} \label{f1}
\end{figure}

Indeed,  skyrmion crystals arise from a  superposition of three  helices with ordering wave vectors $\bm{Q}_{\mu}$~\cite{PhysRevLett.108.017206}:
\begin{equation}\label{eq2}
S_{xy}(\bm{r})=\frac{I_{xy}}{C}\sum_{\mu=1}^3\sin[\bm{Q}_{\mu} \cdot \bm{r}+\theta_{\mu}(l)]\bm{e}_{\mu},
\end{equation}
\begin{equation}\label{eq3}
S_z(\bm{r})=\frac{1}{C}\left[I_{z}\sum_{\mu=1}^3\cos[\bm{Q}_{\mu} \cdot \bm{r}+\theta_{\mu}(l)]+S^0_{z}\right],
\end{equation}
where $I_{xy}, I_z>0$, $S^0_{z}$ is the uniform magnetization induced by ${\bm H}$, $C (\bm{r})=\sqrt{S_x^2+S_y^2+S_z^2}$ and $\bm{e}_{\mu}$ are unit vectors in the $ab$ plane, satisfying $\sum_{\mu=1}^3\bm{e}_{\mu}=0$.  The layer index $l$ is the third component of the verctor ${\bm r}$. For each helix, the spin rotates in a plane parallel to ${\bm H}$.  The angle between the rotation plane and $\bm{Q}_{\mu}$ is arbitrary because of the $U(1)$ symmetry of ${\cal H}$. This freedom implies that skyrmions can have arbitrary helicity $\gamma$~\cite{nagaosa_topological_2013} : $\gamma=0$ for $\bm{e}_{\mu} \parallel \bm{Q}_{\mu}$  (N\'{e}el skyrmion) while $\gamma=\pi/2$ for $\bm{e}_{\mu} \perp \bm{Q}_{\mu}$ (Bloch skyrmion). 

The condition $S_z=-1$ and $S_{xy}=0$, or $\cos(\bm{Q}_{\mu} \cdot \bm{r}+\theta_{\mu})=-1$  is fulfilled at the center of each skyrmion, implying that~\cite{Petrova2011}
\begin{equation}
3 l Q_z+\sum_{\mu=1}^3 \theta_{\mu} (l) = \pi + 2 n \pi, 
\label{constr}
\end{equation}
where $n$ is an arbitrary integer. A translation of the SC in the $ab$ plane is obtained by shifting the phases  $\theta_{\mu}(l)$, subjected to the constraint \eqref{constr}. The choice of $\theta_{\mu}(l)$ corresponds to the different ways of stacking the pancake skyrmions along the $c$-axis, where $l$ is the layer index. For $ABAB\cdots$ stacking ($Q_z=\pi$), we can choose $\theta_{\mu}(l=0)=\pi$, and $\theta_2(l=1)=\theta_3(l=1)=-\theta_1(l=1)/2=2\pi/3$. For $ABCABC\cdots$ stacking ($Q_z=2\pi/3$), we have $\theta_\mu(l)=\pi$~\cite{SM}.

For the Hamiltonian parameters under consideration, the three ordering wave vectors  are $\bm{Q}_1=(Q_{ab},\ 0,\ Q_z)$, $\bm{Q}_2=(-Q_{ab}/2,\ \sqrt{3}Q_{ab}/2,\ Q_z)$, $\bm{Q}_3=(-Q_{ab}/2,\ -\sqrt{3}Q_{ab}/2,\ Q_z)$. 
According to \eqref{constr}, $\theta_{\mu}(l)$ changes with  $l$ if  $Q_z \neq 2 n \pi/3$,  implying the generation of  higher harmonics $m Q_z$ ($m$ is an integer) with a resulting exchange energy cost. In other words,  skyrmion crystals are more stable for $Q_z=0$ or $Q_z=2\pi/3$. Indeed, a single-${\bm{Q}}$ conical state is unstable towards the generation of a second ${\bm{Q}}$-component if~\cite{PhysRevB.93.184413}:
 \begin{equation}
 {\bm{Q}}_1 + {\bm{Q}}_2 + {\bm{Q}}_3=0.
 \label{cond0}
 \end{equation}
 This condition  is  naturally fulfilled in $C_6$ invariant 2D lattices and it still holds for our 3D lattice if $3Q_z=2n\pi$. In general~\cite{SM}, the critical value of $A$ that renders the single-${\bm{Q}}$ conical state unstable is:
\begin{equation}
A_c =   J\left( \bm{Q}_1 + \bm{Q}_2 \right) -  J \left( {{\bm{Q}_3}} \right).
\label{cond}
\end{equation}  
Given that only collinear orderings can survive for large enough $A$, Eq.~\eqref{cond} suggests that  $A_c$ should be significantly smaller than the typical value of the  exchange interaction to guarantee the existence of non-collinear 
multi-${\bm{Q}}$ phases. We will see below that this simple analysis is consistent with Monte Carlo simulations of ${\cal H}$ based on the standard Metropolis algorithm~\cite{SM}. The results presented here are obtained on finite lattices  of $36\times36\times36$ spins with periodic boundary conditions.

\begin{figure}[b]
\psfig{figure=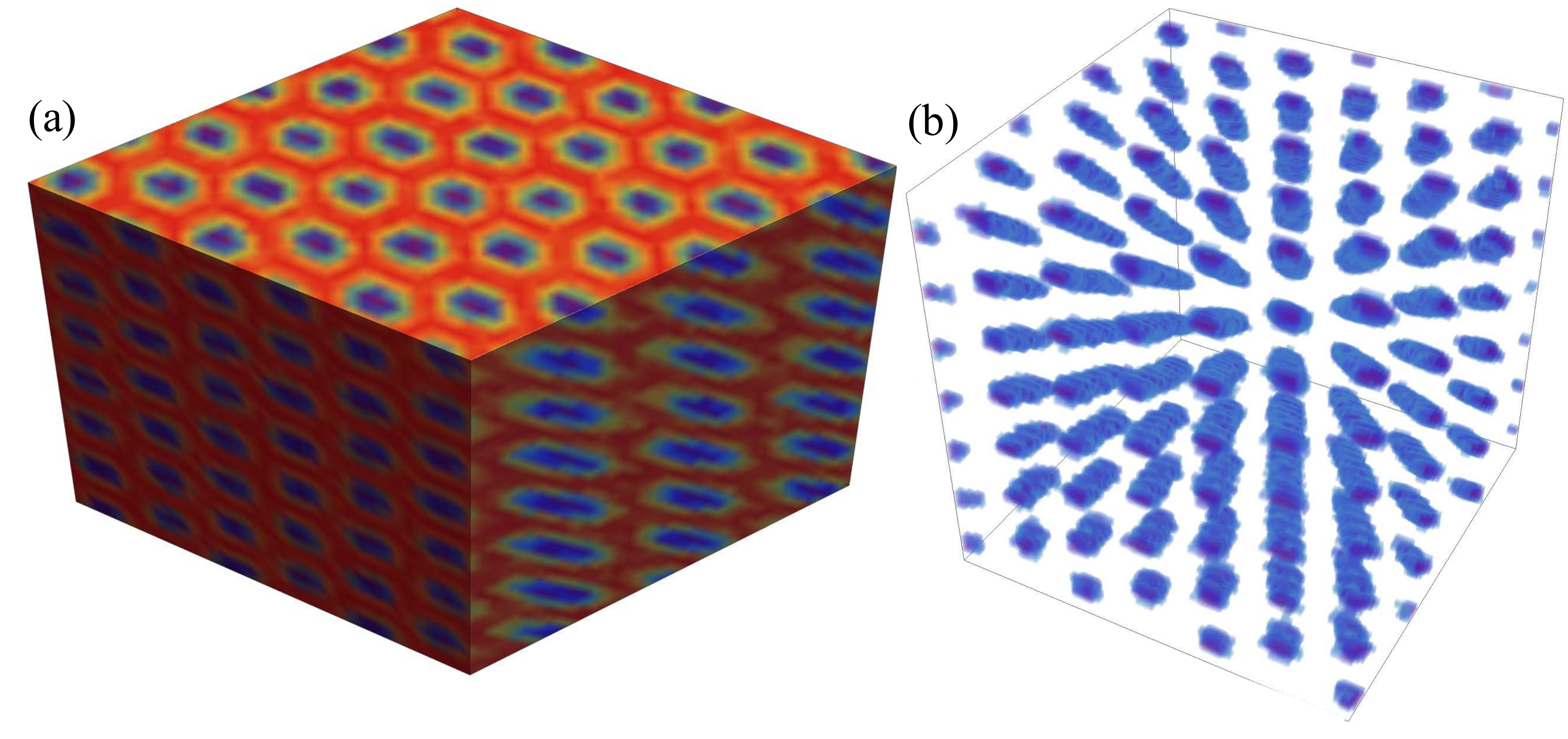,width=\columnwidth}
\caption{(color online) (a) Skyrmion spin configuration at the surface of the simulation box for the titled skyrmion line crystal phase in Fig. \ref{f1}~(b) ($Q_z=2\pi/5$). The color represents the $z$ spin component. (b) Arrangement of skyrmion cores (blue) defined by $S_z<-0.4$. 
} \label{f2}
\end{figure}

We first consider the case of NN FM interlayer exchange $J_1^c<0$ (uniform along the $c$-axis) and anisotropy $A = 0.5 |J_1|$. The resulting phase diagram is shown in Fig.~\ref{f1}~(a). Similarly to the 2D case,~\cite{leonov_multiply_2015,Lin2016a,PhysRevB.93.184413} a vertical spiral (VS) phase (polarization plane parallel to ${\bm H}$) appears in the low field and low temperature region of the phase diagram. The propagation wave vectors are ${\bm Q}_1=(Q_x,\ 0,\ 0)$ and the other two vectors obtained by rotations of $\pm 120^{\circ}$ about the $c$-axis.  A triangular crystal of vertical (parallel to ${\bm H}$) skyrmion lines  is stabilized  below the saturation field. The  SC occupies a large region of the phase diagram because Eq.~\eqref{cond0} is fulfilled ($m Q_z =0 $ implying no exchange energy cost due to generation of higher harmonics).  Indeed, the phase diagram is quite similar to its 2D counterpart,~\cite{leonov_multiply_2015,Lin2016a,PhysRevB.93.184413} except for the appearance of a low-field  collinear modulated (CM) phase, similar to a spin density wave, right next to the paramagnetic (PM) state~\footnote{This phase also appears in a mean field treatment of the problem.}.

\begin{figure}[t]
\psfig{figure=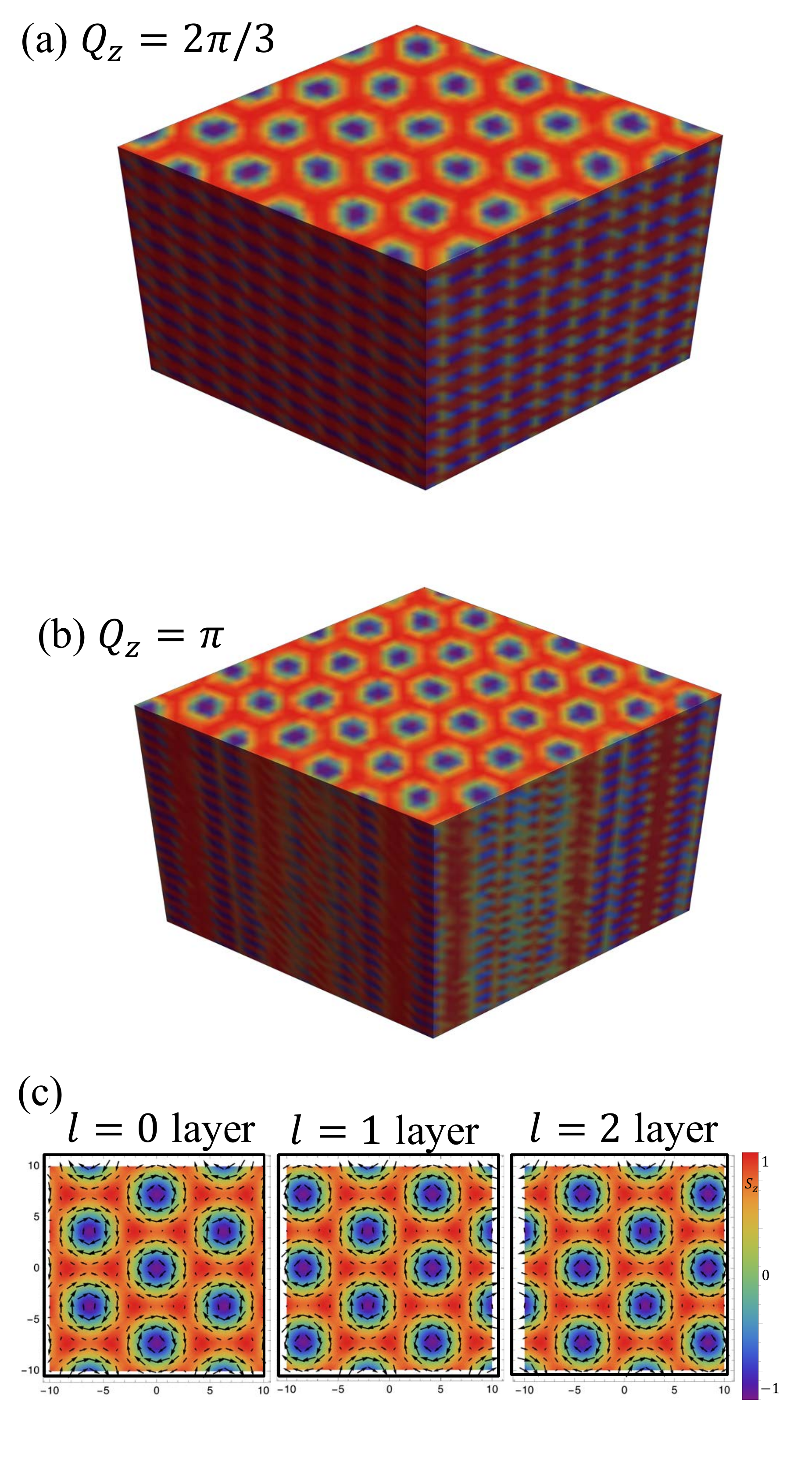,width=\columnwidth}
\caption{(color online) Same as Fig.~\ref{f2} but with (a) $Q_z=2\pi/3$ and (b) $Q_z=\pi$. (c) Spin configuration at different layers are described by  Eqs.~\eqref{eq2}, \eqref{eq3} and \eqref{eqac1}, where $\theta_i(l)$ is chosen to produce the FCC-PSC, corresponding to $ABCABC\cdots$ stacking with $Q_z=2\pi/3$.
} \label{f3}
\end{figure}

We next consider a frustrated interlayer interaction ($J_1^c<0$ and NNN inter-layer exchange $J_2^c>0$). The resulting smooth modulation along the $c$-axis, $Q_z=2\pi/5$ [$J_2^c=-J_1^c/4\cos(Q_z)$], violates the
the condition \eqref{cond0}.
The low-field and low-$T$ phase is still a VS, while a CM phase with the same ordering wave vector appears just below $T_N$. A pancake skyrmion crystal (PSC) is still stabilized for small $J_1^c$ and intermediate magnetic field values. However,  the size of this skyrmion phase is significantly reduced [see Figs.~\ref{f1}~(a) and (b)], as a consequence of  the deviation from the condition \eqref{cond0}.  The skyrmion centers are still smoothly connected in neighboring layers for a small $Q_z$. However,  as shown in Fig.~\ref{f2}~(b), the new $c$-axis modulation has the effect of tilting the  skyrmion lines \emph{away} from the field axis, leading to a tilted skyrmion line crystal (TSLC). 
The tilting   angle $\alpha$ must be compatible with the period ${\tilde c} = 2 \pi/Q_z$ along the vertical direction.
The minimum angle $\alpha$ that satisfies this condition is obtained  by tiling the skyrmion lines along a direction (e.g. [010]) connecting nearest-neighbor (NN) skyrmions (higher values of $\alpha$ are penalized by the inter-layer exchange because they  increase the amplitude  of higher harmonics $mQ_z$). Given the lattice constant of the SC in one layer is ${\tilde a} =4\pi/\sqrt{3}Q_{ab}$,
we get $\tan{\alpha}= {\tilde a}/{\tilde c}=2 Q_z/ \sqrt{3} Q_{ab}$ for the optimal tilting angle.
If the skyrmion lines are tilted in the  [010] direction,  a skyrmion center of the $l$-th layer is located at $R_0(l)=(0, l \tan\alpha)$. According to Eqs.~\eqref{eq2} and \eqref{eq3}, the phases $\theta_{\mu} (l)$ result  from the condition
\begin{equation}
({Q_{\mu,y}}\tan\alpha  + {Q_z})l + {\theta _{\mu}}\left( l \right) = 
3{Q_z}l + {\theta _1}\left( l \right) + {\theta _2}\left( l \right) + {\theta _3}\left( l \right) = \pi, 
\end{equation}
which describes the configuration shown in Fig.~\ref{f2}.
Upon further increasing $H$, the SC undergoes a transition into a double-$Q$ conical (D$Q$C) phase with the spins canted  toward the  field direction and the transverse components rotating with two different propagation wave vectors. 

For $Q_z=2\pi/3$ (frustrated exchange interaction along the $c$-axis), the projection of pancake skyrmion on the adjacent layers lies at the center of the triangle formed by NN skyrmions on those layers. This situation is energetically favored because the magnetization at the skyrmion core is opposite to the magnetization at the center of the triangle formed by three skyrmions (interstitial). The $ABCABC\cdots$  stacking (3-layer period consistent with $Q_z=2\pi/3$) shown in Fig.~\ref{f3}~(a) corresponds  to a  face centered cubic (FCC) pancake skyrmion crystal (FCC-PSC).   Given that each skyrmion center satisfies the condition $S_z(R_i)=-1$, the skyrmion position $\bm{R}_i(l)=[ x(l) ,\ y(l) ]$ at $l$-th layer is determined from
\begin{align}\label{eqac1}
{{\bm{Q}}_{\nu,ab}} \cdot \bm{R}_i + {Q_z}l + {\theta _{\nu}}\left( l \right) = (2n_{\nu} + 1)\pi, 
\end{align}
where $\nu=1,2$ and $n_{\nu}$  are integer numbers. Here $\theta_{\mu}(l)$ satisfies Eq.~\eqref{constr}.  We now fix one pancake skyrmion center at $ {\bm R}_0(l=0)=(0, 0)$, i.e., $\theta_{\mu}(l=0)=\pi$ for $\mu =1,2,3$. For $l=1$, a pancake skyrmion center is located at ${ \bm R}_0(l=1)=({\tilde a}/\sqrt{3},\ 0)$.  The sequence shown in the three consecutive panels of Fig.~\ref{f3}(c) is obtained by selecting $\theta_{\mu}(l)=\pi$ for all  layers.  Figure~\ref{f3}~(c) shows the resulting spin configurations for the $l=0$, $l=1$ and $l=2$ layers. As shown in Fig.~\ref{f1}(c) for $J^c_{2}=0.5J^c_{1}=-0.5 J_1$, the FCC-PSC occupies a wide region of the phase diagram because Eq.~\eqref{cond0} is fulfilled.

The strongest deviation from Eq.~\eqref{cond0} corresponds to $Q_z=\pi$. Figure~\ref{f1}~(d) shows the  $H$-$T$ phase diagram obtained for $J^c_{1}=-0.2J_1$ and $J^c_2=0$. 
The resulting spin configuration is the $ABAB\cdots$ stacking of pancake skyrmions shown in Fig.~\ref{f3}~(b) corresponding to a hexagonal closed-packed  (HCP) PSC. To reproduce the HCP-PSC with the ansatz of \eqref{eqac1}, we can 
choose the pancake skyrmion positions at layers $l=0$ and $l=1$ to be the same as for the FCC-PSC and repeat the pattern for even and odd layer:  $\theta_{\mu}(l=0)=\pi$ and $\theta_2(l=1)=\theta_3(l=1)=-\theta_1(l=1)/2=2\pi/3$.
The HCP-PSC is suppressed for large enough values of the AFM interlayer exchange~\cite{SM}.

\begin{figure}[t]
\psfig{figure=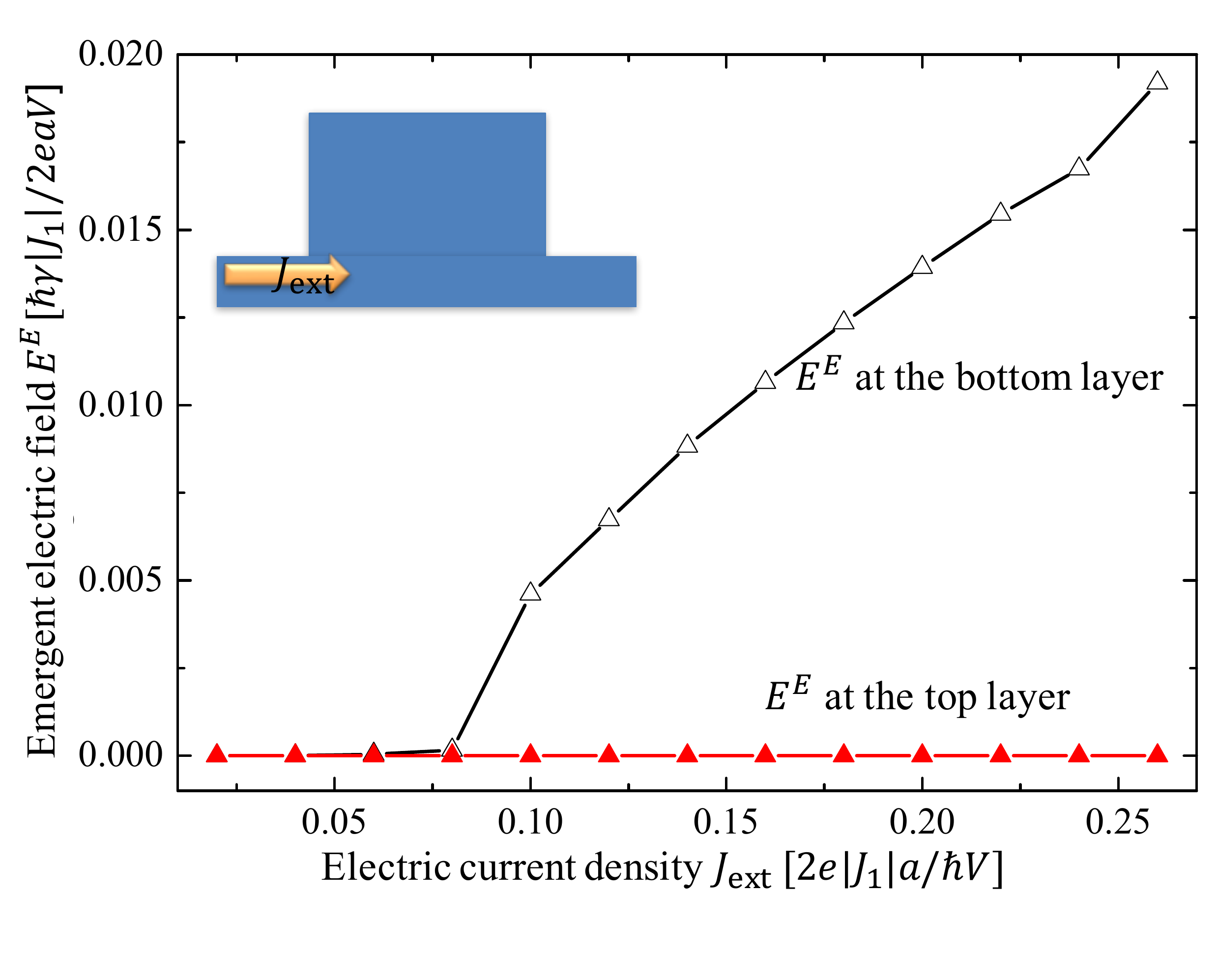,width=\columnwidth}
\caption{(color online)  Emergent longitudinal electric field as a function of current $J_{\mathrm{ext}}$ applied at the bottom layer. The Hamiltonian parameters are the same as in Fig.~\ref{f1}~(d) with $H=0.64 |J_1|$. Here $a$ and $V$ in the units are the lattice parameter and volume of a unit cell, respectively.
} \label{f4}
\end{figure}

A moderate easy-axis anisotropy is essential for stabilizing the above-described skyrmion crystals. However, a strong deviation from Eq.~\eqref{cond0} will drastically reduce the stability of any multi-$\bm{Q}$ ordering~\cite{SM}.

The skyrmion crystals discussed so far arise from a superposition of three helices. To demonstrate the particle nature of pancake skyrmions, we inject an inhomogeneous current. The spin dynamics  obeys  the Landau-Lifshitz-Gilbert equation
\begin{equation}\label{eqdynamics}
{\partial _t}{\bm{S}} = \frac{\hbar\gamma}{2e}({{\bm{I}_{\mathrm{ext}}} }\cdot\nabla) {\bm{S}} - \gamma {\bm{S}} \times {{\bm{H}}_{\rm{eff}}} + \alpha {\bm{S}}\times  {\partial _t}{\bm{S}} ,
\end{equation}
where $\gamma$ is the gyromagnetic ratio, $\alpha$ is the Gilbert damping constant, $\bm{H}_{\rm{eff}}\equiv-\delta \mathcal{H}/\delta{\bm{S}}$ is the effective magnetic field and $\bm{I}_{\mathrm{ext}}=\bm{J}_{\mathrm{ext}}\delta_{l,0}$ is the spin polarized current injected at the bottom layer $l=0$. The current flows in the $[100]$ direction of the hexagonal spin lattice. The emergent electric field induced by the skyrmion motion, $\bm{E}^E=\hbar\bm{S}\cdot(\nabla\bm{S}\times\partial_t\bm{S})/(2e)$,  is proportional to the skyrmion velocity. The resulting $\bm{E}^E$ for the HCP-PSC at the bottom and top surfaces are shown in Fig. \ref{f4}. Skyrmions remain pinned by the discrete spin lattice for  small currents. Pancake skyrmions in the bottom layer start moving
(nonzero $E^E$) when the current reaches a threshold value,   while the skyrmions at the top surface remain at rest. We have not observed an intermediate region, where skyrmions at the top surface are dragged by the motion of skyrmion in the bottom layer. Such an intermediate region has been observed in chiral skyrmion phase \cite{PhysRevB.93.060401}. The absence of the intermediate region in the HCP pancake skyrmion lattice is probably due to the fact that the skyrmion pinning force is stronger than the coupling between skyrmions in adjacent layers. The weak interlayer coupling allows pancake skyrmions to decouple from other layers. The particle nature of the pancake skyrmion can also be seen from a metastable configuration of pancake skyrmions obtained by simulated annealing~\cite{SM}.

In summary, we have demonstrated that different 3D skyrmion crystals can be stabilized in centro-symmetric magnets by tuning the ratio between competing inter-layer exchange interactions. Pancake skyrmions stack uniformly along the $c$-axis (magnetic field direction)  for FM  interlayer coupling $(Q_z=0)$, leading to a triangular lattice of skyrmion lines. A small $Q_z$ has the effect of tilting the skyrmion lines away from the  $c$-axis. 
Much larger values of $Q_z$ lead to an HCP-PSC for $Q_z=\pi$ and an FCC-PSC for $Q_z=2\pi/3$. As expected from the analysis that lead to Eq.~\eqref{cond0}, the skyrmion crystals are more stable for $Q_z=0$ and $Q_z=2\pi/3$. These novel spin configurations  can be realized in rare-earth magnets with Ruderman-Kittel-Kasuya-Yosida interaction and  moderate easy-axis anisotropy \cite{JensenBook}, as well as in frustrated magnets~\cite{batista_frustration_2016}. Finally, we note that superconducting pancake vortices of layered superconductors can also stack at a finite angle relative to the magnetic field direction because of the underlying crystal anisotropy \cite{Blatter94}. However, HCP or FCC crystals of pancake vortices have never been observed.

\begin{acknowledgments}
The authors thank Ivar Martin for helpful discussions. Computer resources for numerical calculations were supported by the Institutional Computing Program at LANL. This work was carried out under the auspices of the U.S. DOE contract No. DE-AC52-06NA25396 through the LDRD program.
\end{acknowledgments}

\bibliography{reference}

\newpage
\clearpage
\appendix
 
\begin{figure*}[t]
\psfig{figure=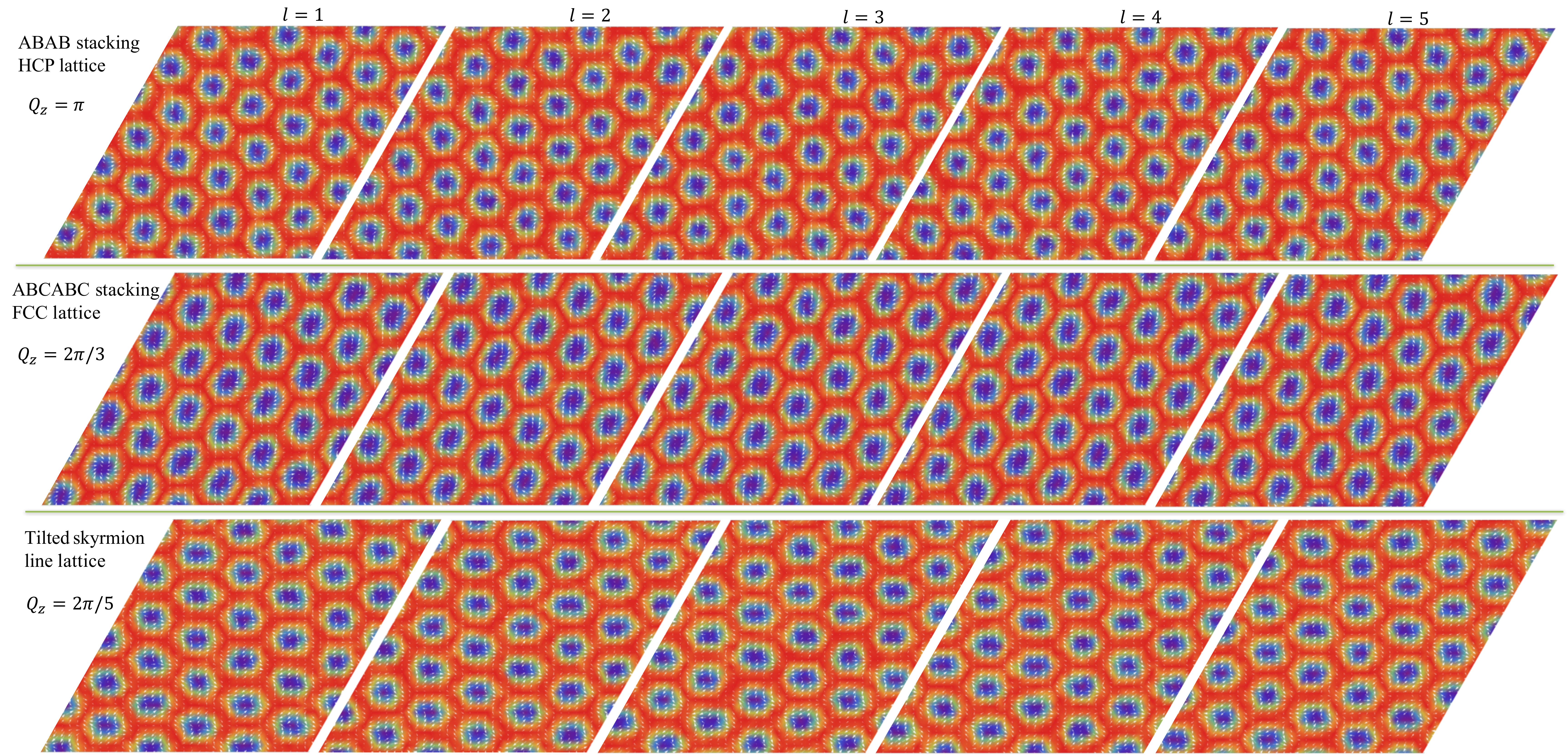,width=18.5cm}
\caption{(color online) Spin configuration for $Q_z=\pi$ (first row), $Q_z=2\pi/3$ (second row), $Q_z=2\pi/5$ (third row) for $l=1,\ 2 ,\ 3 ,\ 4 ,\ 5 $ layers  obtained by simulations. Color represents the $z$ component of the spin and arrows denote the in-plane components. The plots correspond to the results in Figs. \ref{f2} and \ref{f3} in the main text.
} \label{fs1}
\end{figure*}

\section{Configuration of pancake skyrmions at each layer: simulations}\label{apx2}
In Fig.~\ref{fs1}, we show the spin configuration for $Q_z=\pi$ (first row), $Q_z=2\pi/3$ (second row), $Q_z=2\pi/5$ (third row) for $l=1,\ 2 ,\ 3 ,\ 4 ,\ 5 $ layers  obtained by numerical simulations. We note that  the period of the magnetic structures is always the same because it is dictated by the in-plane component of the ordering wave vectors ${\bm Q}_{\mu}$, which remains fixed in our calculations ($Q_{ab}= 2 \pi/5$). In contrast, the diameter of the skyrmion cores varies as a function of magnetic field, anisotropy and $Q_z$. For instance, as it is clear from  Fig.~\ref{fs1}, the skyrmion cores shrink for $Q_z=\pi$ because the inter-layer AFM interaction penalizes the overlap between skyrmion cores (the spins in the skyrmion cores are polarized in the same direction antiparallel to the external field). Similarly, the skyrmion cores shrink as a function of increasing magnetic field. 

\begin{figure}[t]
\psfig{figure=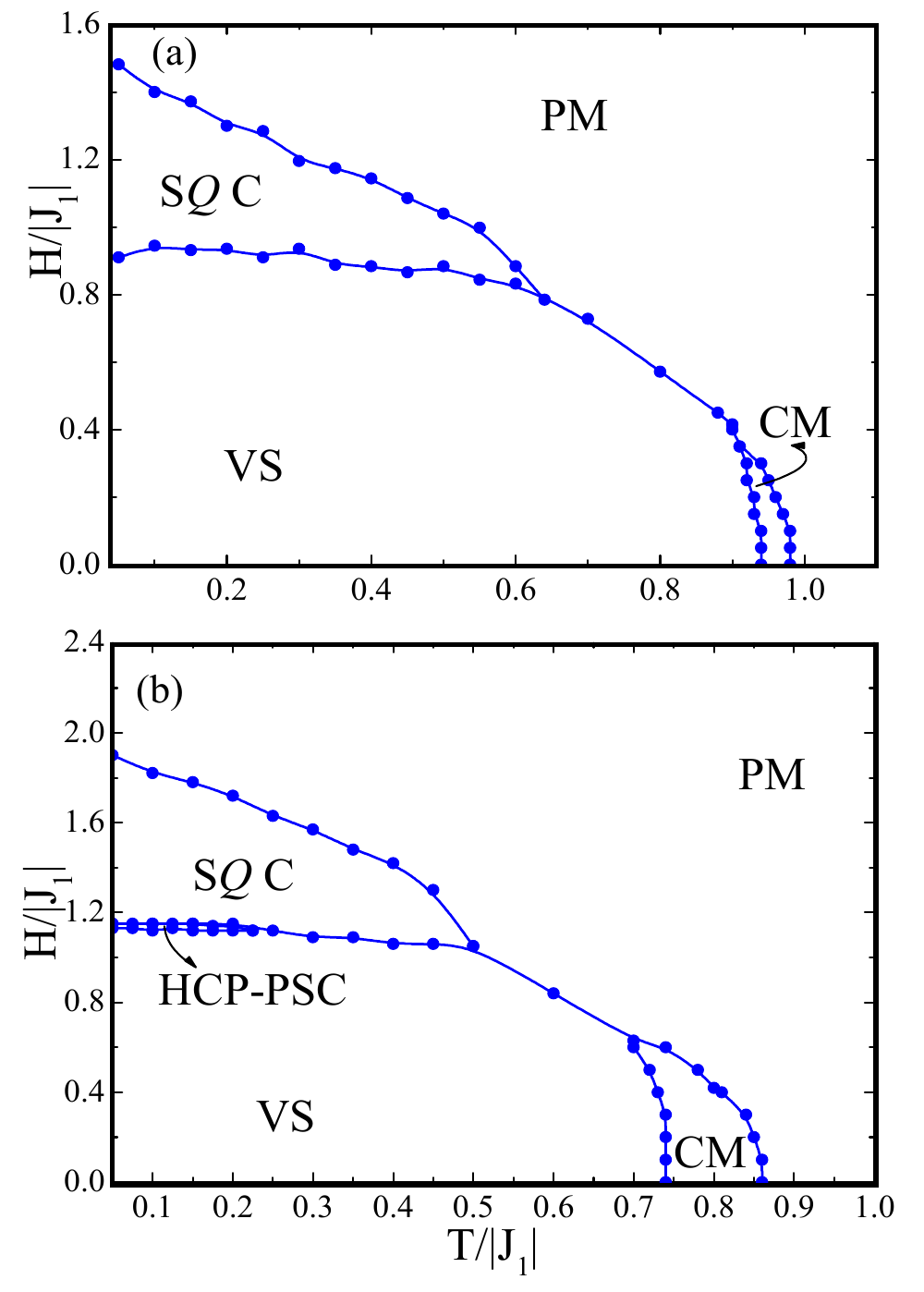, width=7.0cm}
\caption{(color online) Temperature-magnetic field phase diagram for a stronger interlayer coupling. Here (a) $Q_z=2\pi/5$ for $J_{1}^c=1.0J_1$ and take $J_{2}^c=-0.809017J_{1}^c$. (b) $Q_z=\pi$ for $J_1^c=-0.5J_1$ and $J_2^c=0$. The easy axis anisotropy is $A=0.5|J_1|$.
} \label{fs3}
\end{figure}

\section{Stability of the single-${\bf Q}$ conical state}\label{apx1}

\begin{figure}[t]
\psfig{figure=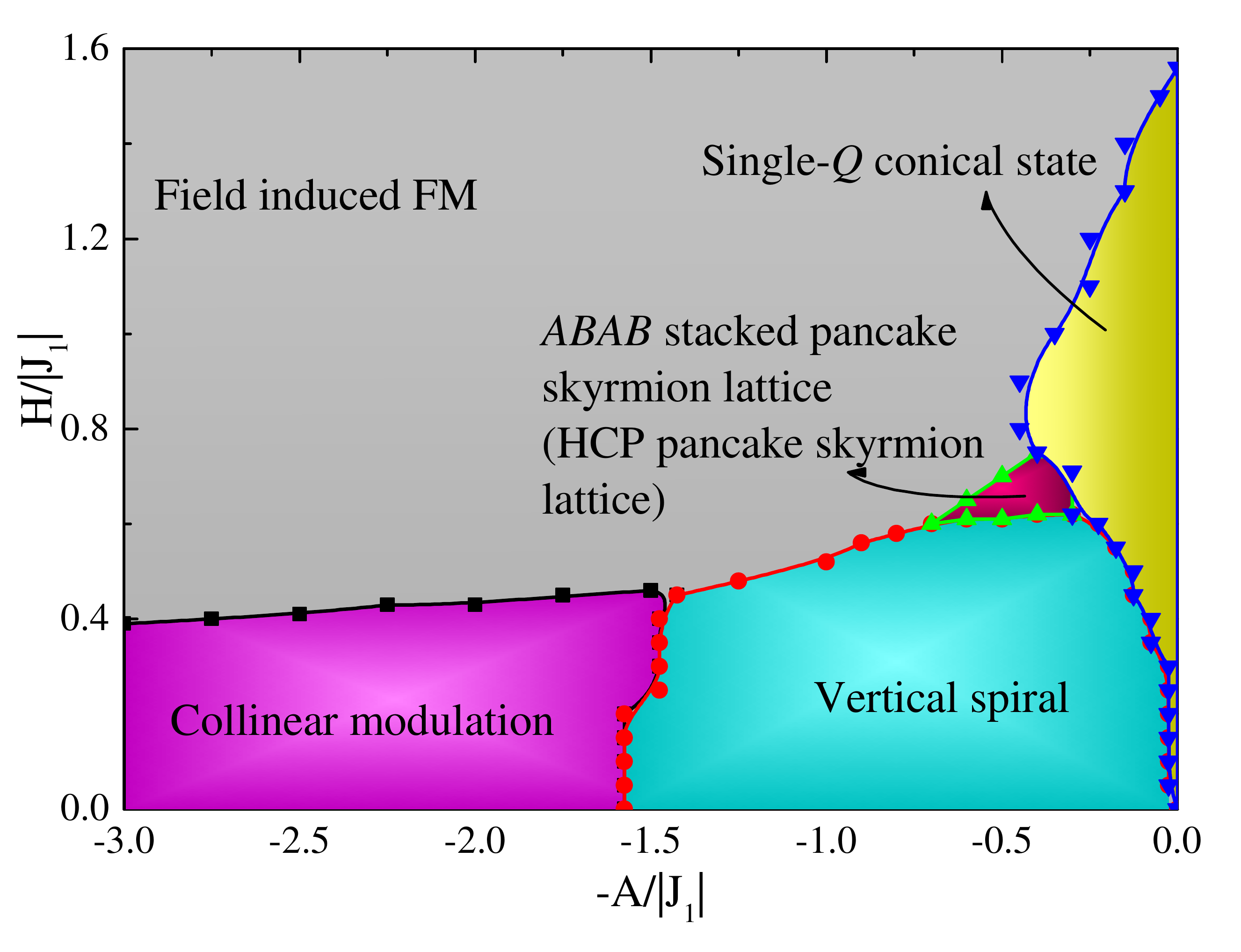,width=\columnwidth}
\caption{(color online) Easy-axis anisotropy-magnetic field phase diagram at $T=0.1|J_1|$. Here $Q_z=\pi$ in the presence of a NN AFM interlayer coupling $J_1^c=-0.2 J_1$ and $J_2^c=0$. } \label{fs4}
\end{figure}

Here we analyze  the stability of the single-${\bm Q}$ solution  in the presence of easy-axis anisotropy and
interlayer exchange interactions $J^c_1$ and $J^c_3$. 
Like for the 2D case,~\cite{leonov_multiply_2015,Lin2016a,PhysRevB.93.184413} the ordered phase is a conical state in absence of anisotropy ($A=0$). At low fields, the easy-axis anisotropy destroys the conical state in favor of the VS phase that is discussed in the main text (see Fig.~\ref{f1} in the main text). However, the VS phase cannot be continuously connected with the fully polarized state. Moreover, a uniform magnetization component is induced at the expense of generating higher harmonics, which are penalized by the exchange interaction. This situation leads to two possible scenarios, which are illustrated 
by the phase diagram of  Fig.~\ref{fs4}. The first and simplest scenario corresponds to a direct strongly first order transition from the  VS state to the fully polarized state. The second scenario includes an intermediate multi-$\bm{Q}$ phase 
between the VS phase and the fully polarized state. A simple criterion for existence of such a multi-$\bm{Q}$ phase can be derived from the following stability analysis of the single-${\bm Q}$ conical phase.

\begin{figure}[b]
	\psfig{figure=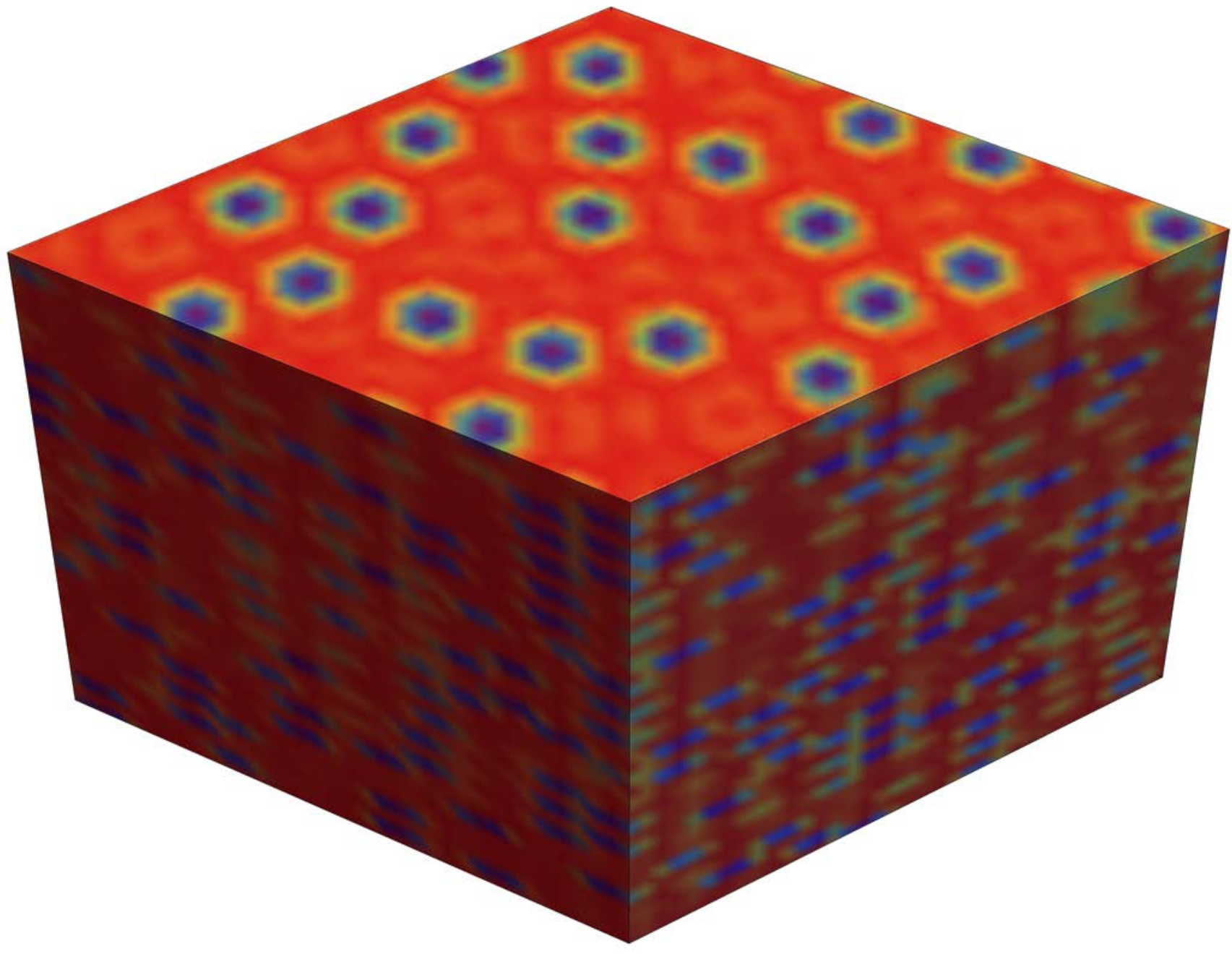,width=\columnwidth}
	\caption{(color online) Metastable pancake skyrmion configuration obtained by simulated annealing.} \label{fs5}
\end{figure}

We start by considering the following (double-$Q$) deformation of the conical state
\begin{align}
{S_x}\left( \bm{r} \right) = \sqrt {{{\sin }^2}\tilde \theta  - {\Delta ^2}} \cos \left( {\bm{Q}_1\cdot\bm{r}} \right) + \Delta \cos \left( {\bm{Q}_2\cdot\bm{r}} \right),
\end{align}
\begin{align}
{S_y}\left( \bm{r} \right) = \sqrt {{{\sin }^2}\tilde \theta  - {\Delta ^2}} \sin \left( {\bm{Q}_1\cdot\bm{r}} \right) - \Delta \sin \left( {\bm{Q}_2\cdot\bm{r}} \right),
\end{align}
\begin{align}
{S_z}\left( \bm{r} \right) = \sqrt {{{\cos }^2}\tilde \theta  - 2\Delta \sqrt {{{\sin }^2}\tilde \theta  - {\Delta ^2}} \cos \left( {\bm{Q}_s\cdot\bm{r}} \right)},
\end{align}
where $\bm{Q}_1$ and $\bm{Q}_2$ are two ordering wave vectors and $\bm{Q}_s=\bm{Q}_1+\bm{Q}_2$. The magnitude of the deformation is parametrized by $\Delta$. Assuming $\Delta\ll 1$. By expanding  $S_z(\bm{r})$
for a small $\Delta$ 
\begin{align}
{S_z}\left( \bm{r} \right) \approx \cos \tilde \theta \left( 1 - x\cos \left( {\bm{Q}_s\cdot\bm{r}} \right) - \frac{1}{2}{x^2}{{\cos }^2}\left( {\bm{Q}_s\cdot\bm{r}} \right)\right) \nonumber\\
-  \cos \tilde \theta\left(\frac{1}{2}{x^3}{{\cos }^3}\left( {\bm{Q}_s\cdot\bm{r}} \right) + \frac{5}{8}{x^4}{{\cos }^4}\left( {\bm{Q}_s\cdot\bm{r}} \right) \right),
\end{align}
with
\begin{align}
x = \frac{{\Delta \sqrt {{{\sin }^2}\tilde \theta  - {\Delta ^2}} }}{{{{\cos }^2}\tilde \theta }},
\end{align}  
and substituting in
\begin{align}
\frac{\cal H }{N} =   \frac{1}{N} \sum_{\bm q} J \left( {\bm q} \right) {\bm S}\left( {\bm q} \right)\cdot {\bm S}
\left(  - {\bm q}  \right)
 - \frac{H}{N} \sum_i S_{i,z} - \frac{A}{N} \sum_i S_{i,z}^2,
\end{align}  
we obtain the energy per spin. Here $N$ is the total number of spins, while $J(\bm{q})$ and $\bm{S}(\bm{q})$ are the Fourier transform of the exchange interaction $J_{ij}$ and the spin $\bm{S}_i$. The result is
\begin{align}
\label{expan}
\frac{{\cal H}}{N} =   {\Delta ^2}J\left( {{\bm{Q}_2}} \right) + \left( {{{\sin }^2}\tilde \theta  - {\Delta ^2}} \right)J\left( {{\bm{Q}_1}} \right)\nonumber\\
 + {\cos ^2}\tilde \theta \left[ J\left( 0 \right) + \frac{1}{2}\left( { - J\left( 0 \right) + J\left( {{\bm{Q}_s}} \right)} \right){x^2}\right] \nonumber\\
 + \frac{{\cos ^2}\tilde \theta}{{32}}\left[\left( { - 13J\left( 0 \right) + 12J\left( {{\bm{Q}_s}} \right) + J\left( {2{\bm{Q}_s}} \right)} \right){x^4} \right] \nonumber\\
- H\cos \tilde \theta \left( {1 - \frac{1}{4}{x^2} - \frac{{15}}{{64}}{x^4}} \right) - A{\cos ^2}\tilde \theta. 
\end{align}
We need to minimize energy with respect to $\tilde{\theta}$ and $\Delta$. 
To zeroth order in $\Delta$, we have
\begin{align}\label{eqaa9}
\cos\tilde\theta  = \frac{H}{{2\left[ {-A + J\left( 0 \right) - J\left( \bm{Q} \right)} \right]}}.
\end{align}
By keeping only terms up to order $\Delta^2$ in the expansion \eqref{expan}, we obtain
\begin{align}\label{eqaa8}
\frac{{\cal H}}{N} =  { - H \cos \tilde \theta - \left[ {A - J\left( 0 \right)} \right]\cos^2 {{ \tilde\theta }} + J\left( \bm{Q} \right)\sin^2 {{ \tilde\theta}}} \nonumber  \\
+ \frac{1}{4}\left[ {-2J\left( 0 \right) + 2J\left( {{\bm{Q}_s}} \right) + H\sec  \tilde\theta } \right]{\Delta ^2}\tan^2  \tilde\theta ,
\end{align}
where $J(\bm{Q}) =J(\bm{Q}_1)=J(\bm{Q}_2)=J(\bm{Q}_3)$ and $\cos \tilde \theta $ is given by \eqref{eqaa9}.
We then obtain 
\begin{align}
\frac{{{\cal H}\left( \Delta  \right) - {\cal H}\left( {\Delta  = 0} \right)}}{N} = \frac{1}{2}\left[ {-A - J\left( \bm{Q} \right) + J\left( {{\bm{Q}_s}} \right)} \right]{\Delta ^2}\tan^2  \tilde\theta.
\end{align}
The conical phase is unstable if this quantity is negative.
 When $Q_z=0$ or $Q_z=2\pi/3$, we have $J(\bm{Q})=J(\bm{Q}_s)$, and $\mathcal{H}(\Delta)-\mathcal{H}(\Delta=0)<0$, implying that  the single-$\bm{Q}$ conical phase is unstable for an infinitesimal $A>0$  (easy-axis anisotropy). For other values of $Q_z$, we have $J(\bm{Q})<J(\bm{Q}_s)$, implying that a threshold value
 of $A_c$ is required to render the single-$\bm{Q}$ conical phase unstable:
\begin{align}
-A_c + J\left( \bm{Q}_1+ \bm{Q}_2\right) - J\left( {{\bm{Q}}} \right) = 0.
\end{align}  
This simple analysis explains why the multi-$\bm{Q}$ high-field phase  is strongly reduced or even completely suppressed for $Q_z$ values which are far from $0$ or $2\pi/3$ [see Figs.~\ref{f1}(b) and (d) in the main text, as well as Fig.~\ref{fs4}]. 
The amplitude of the second modulation with a wavevector $\bm{Q}_2$ is determined by the ${\cal O}(\Delta^4)$ term of
the expansion \eqref{expan}. 

\section{Numerical details}\label{apx3}

Our Monte Carlo (MC) simulations are based on the standard Metropolis algorithm. Starting from a random spin configuration, the system is annealed with $4\times10^6$ MC sweeps (MCS), followed by  $5\times10^6$ MCS to reach equilibrium and  $5\times10^6$ MCS to measure the relevant quantities. Periodic boundary conditions are imposed in all directions. Most simulations are done on a cluster of $36\times 36\times 36$ sites. Additional simulations on larger lattices ($45\times 45\times 45$) were performed to check for finite size effects. The phase boundary is determined by analyzing the spin structure factor, spin susceptibility and specific heat as a function of magnetic field and temperature. We have verified that the phase boundary of the skyrmion crystal phases is the same for lattices of $36\times 36\times 36$ and $45\times 45\times 45$. The period of the obtained magnetic structures (linear size of the magnetic unit cell) is  $2 \pi/ |{\bm Q}_{\mu, ab}| =5$. This period is always the same in our simulations because we are fixing the ratio $J_3/J1$. A  linear lattice size of $L=45$ sites corresponds to nine magnetic unit cells. According to our simulations, this ratio (nearly one order of magnitude) is large enough to suppress undesirable finite size effects.

The Landau-Lifshitz-Gilbert equation is  solved by an explicit numerical scheme developed in Ref. \onlinecite{Serpico01}. We use periodic boundary condition in the $x$ and $y$ direction and open boundary condition in the $c$ direction. The system is annealed to reach the ground state before applying a current in the bottom layer.

\section{Results for strong interlayer competing interactions}\label{apx4}

We increase $J_{1}^c$ so that $J(\bm{q})$ is dominant by the interaction along the $c$ axis. For $Q_z=2\pi/5$, we use $J_{1}^c=1.0J_1$ and take $J_{2}^c=-0.809017J_{1}^c$. In this case, the skyrmion lattice disappears completely, as shown in Fig.~\ref{fs3}~(a). The double-$Q$ conical phase in Fig.~\ref{f1} (b) in the main text is replaced by the single-$Q$ conical (S$Q$C) phase. For $Q_z=\pi$ (AFM interlayer coupling), we take $J_{1}^c=-0.5J_1$ and $J_2^c=0$. In this case, the HCP-PSC phase is extremely narrow [see Fig.~\ref{fs3}~(b)]. For general values of $Q_z$, the SC phase is replaced by a single-$Q$ solution for strong enough competing interlayer coupling. As discussed above,  the strong interlayer coupling does not destroy the PSC for $Q_z=0$ and $Q_z=2\pi/3$.

\section{$H$-$A$ phase diagram}\label{apx5}
Figure~\ref{fs4} shows the low-temperature ($T=0.1 |J_1|$) $A-H$ phase diagram for $Q_z=\pi$. As anticipated, the only ordered phase for $A \to 0$ is a  single-$\bm{Q}$ conical phase.
Unlike the 2D case,~\cite{leonov_multiply_2015} a double-$\bm{Q}$ phase does not appear at high fields upon increasing $A$ from zero.
This is a direct consequence of the deviation from the condition Eq. \eqref{cond0} in the main text. This deviation also explains the drastic reduction of the size of the HCP-PSC, which is consistent with the detailed analysis in  Appendix \ref{apx1} and \ref{apx4}. As expected, a sufficiently strong anisotropy destroys the skyrmion phase in favor of a collinear phase, which is modulated along one direction. In the 2D limit, the collinear high field phase corresponds to a triangular bubble crystal.~\cite{PhysRevB.93.184413} Once again, this triple-$\bm{Q}$ collinear phase disappears in our 3D model with $Q_z=\pi$, because of the violation of  Eq.~\eqref{cond0} in the main text.

\section{Metastable pancake skyrmions configuration}\label{apx6}
In Fig. \ref{fs5}, we show a metastable pancake skyrmion configuration obtained by simulated annealing. In this case, the pancake skyrmions are well separated in space and behave as particles.

\end{document}